\documentstyle[psfrag,graphicx,aps,twocolumn]{revtex}
\def\be{\begin{eqnarray}}
\def\ee{\end{eqnarray}}

\def\roughly#1{\mathrel{\raise.3ex\hbox{$#1$\kern-.75em%
\lower1ex\hbox{$\sim$}}}}

\def\Tr{{\rm Tr}\,}

\begin{document}

\renewcommand{\thefootnote}{\arabic{footnote}}
\setcounter{footnote}{0}

\vskip 0.4cm

\title{\LARGE\bf L\'{e}vy Matrices and Financial Covariances}

\author{
Zdzis\l{}aw Burda$^{a,b}$\thanks{E-mail: burda@physik.uni-bielefeld.de}, 
Jerzy Jurkiewicz$^{a}$\thanks{E-mail: jjurkiew@th.if.uj.edu.pl},
Maciej A. Nowak$^{a}$\thanks{E-mail: nowak@th.if.uj.edu.pl },\\
Gabor Papp$^{c,d}$\thanks{E-mail: pg@ludens.elte.hu}
and Ismail Zahed$^{c}$\thanks{E-mail: zahed@zahed.physics.sunysb.edu}}

\address{
$^a${\it M. Smoluchowski Institute of Physics,
Jagellonian University, Cracow, Poland} \\
$^b${\it Fakult\"at f\"ur Physik, Universit\"at Bielefeld
P.O.Box 100131, D-33501 Bielefeld, Germany} \\
$^c${\it Department of Physics and Astronomy,
SUNY-Stony-Brook, NY 11794  U.\,S.\,A.} \\
%$^d${\it Physics Department, Brookhaven National Laboratory,
%Upton, NY 11973, U.\,S.\,A.} \\
$^d${\it HAS Research Group for Theoretical Physics, 
E\"otv\"os University, Budapest, H-1518 Hungary}}

\date{\today}
\maketitle
\begin{abstract}
In a given market, financial covariances capture the intra-stock
correlations and can be used to address statistically the bulk
nature of the market as a complex system. We provide a statistical
analysis of three  SP500 covariances with evidence for raw tail 
distributions. We study the stability of these tails against
reshuffling for the SP500 data and show that the
covariance with the strongest tails is robust, with a spectral
density in remarkable agreement with random L\'{e}vy matrix
theory. We study the inverse participation ratio for the three
covariances. The strong localization observed at both ends of
the spectral density is analogous to the localization exhibited
in the random L\'{e}vy matrix ensemble. We discuss 
two competitive mechanisms responsible for the occurrence 
of an extensive and delocalized eigenvalue at the edge 
of the spectrum: (a) the L\'{e}vy character of the entries of the
correlation matrix and (b) a  sort of off-diagonal order induced by
underlying inter-stock correlations. (b) can be destroyed by reshuffling,
while (a) cannot. We show that the stocks with the largest 
scattering are the least susceptible to correlations, 
and likely candidates for the localized states. 
We introduce a simple model for price fluctuations
which captures behavior of the SP500 covariances.
It may be of importance for assets diversification.

\end{abstract}

%\renewcommand{\thefootnote}{\#\arabic{footnote}}
%\setcounter{footnote}{0}

%%%%%%%%%%%%%%%%%%%%%%%%%%%%%%%%%%%%%%%%%%%%%%%%%%%%%%%%%%%%%%%%%%%%%%%%%%%
%%%%%%%%%%%%%%%%%%%%%%%%%% Introduction  %%%%%%%%%%%%%%%%%%%%%%%%%%%%%%%%%%
%%%%%%%%%%%%%%%%%%%%%%%%%%%%%%%%%%%%%%%%%%%%%%%%%%%%%%%%%%%%%%%%%%%%%%%%%%%

\section{Introduction}

A number of phenomena in nature are characterized by a 
coexistence of different scales, usually described by 
power law distributions. This is the case of most phase 
transitions where at the critical point the correlation 
functions are scale invariant, as well as most fluid phases
in highly developed turbulence where the velocity fluctuations are
sensitive to a variety of eddies. Power law distributions are also encountered 
in a number of biophysical settings as well as financial markets~\cite{BOOK}.

Stable random phenomena with power law behaviors are usually described
by L\'{e}vy distributions, a consequence of the central limit theorem
for scale-free processes. The simplest example is a random walk 
with a power law distribution for single independent steps, 
where the relative probabilities at different times
are scale free. These phenomena lead to anomalous diffusion and
intermittency as encountered in charge transport in amorphous
semiconductors, moving interfaces in porous media, spin glasses,
turbulence~\cite{BOOKLEV} and phase changes in chiral QCD~\cite{QCD}. 

Recently, it was pointed out that current market covariances are
gaussian noise driven with possible consequences for the assessment 
of correlations in portfolio evolution and optimization~\cite{RAN}. 
In particular, it was shown that the lower part of
eigenvalue distribution of 
the SP500 covariance matrix constructed from the daily returns 
normalized by the local volatility, is Gaussian noise dominated. In 
this paper we confirm some of these observations, but suggest that
an alternative covariance constructed from the daily returns normalized
to the initial price displays L\'{e}vy noise throughout the spectrum.
The latter is more robust against reshuffling and certainly requires
a random L\'{e}vy matrix description. This observation is overall
consistent with a recent observation we made in the context of free 
random L\'{e}vy matrices~\cite{USII.5}.

The outline of the paper is as follows: in section 2, we 
introduce the concept of financial correlation matrices,
and empirically analyze their statistical content. 
We show that all covariances display power law tails, albeit with
different indices. In section 3 we discuss the issue 
of inter-stock correlations and we define reshuffling
of the price series and investigate its effects on
the covariance matrices. We find that the covariance with
the largest tails is the least sensitive to this process.
This is discussed in section 4, where we analyze the 
corresponding spectral densities and show that the results of
random L\'{e}vy matrices apply remarkably well to the covariance
that is stable under reshuffling. In section 5, we analyze the
bulk eigenvector content of the all covariances and parallel
them with the results of reshuffling and L\'{e}vy random matrix
theory. The larger the tails, the stronger the localization 
seen in the participation ratios and the stock scattering.
In section 6, we formulate a simple model 
of price fluctuations, which reproduces most of the
experimentally observed features of the 
SP500 covariances.  Our conclusions are in section 7.

\section{Financial Covariances}

One of the central problems in financial investment is the assessment
of risk. Standard lore suggests that risk can be reduced through
assets diversification, with Markowitz's portfolio analysis as one
of the corner-stones in assets allocation and diversification~\cite{BOOK}.
The key to Markowitz's analysis is the concept of a covariance 
matrix. In this section we define and empirically analyze the 
distribution of entries and also correlations as captured 
in certain SP500 covariance matrices, with comparison to 
results from random L\'{e}vy matrices.
Throughout, we will use price return data from the SP500
daily quotations of $N=406$ stocks over the period of $T+1=1309$ days 
from 01.01.1991 till 06.03.1996 (ignoring dividends). 

\subsection{SP500 covariances}

Consider the covariance matrix constructed from the raw returns
normalized by the initial price:
\be
{\bf C}_{ij}= 
\frac 1T \,\sum_{t=1}^T \, M_{ti} M_{tj} =
\frac 1T \,\sum_{t=1}^T \, 
\frac{m_{ti}}{x_{0i}} \frac{m_{tj}}{x_{0j}}\,\,.
\label{ad1}
\ee
The raw returns $m_{ti}$ of stock $i$ (out of a total of $N$) at 
time $t$, labeled by an integer ($t=1,\dots T$) 
are evaluated at fixed time intervals in a given market as
\be
m_{ti}= {\delta x}_i(t) -{\delta \bar{x}}_i
\label{1}
\ee
where: $\delta x_i(t) = x_i(t+1) - x_i(t)$. 
The mean $\delta {\bar{x}}_i=\sum_t \,{\delta x}_{i}(t)/T$
is subtracted\footnote{In fact, all results presented
in the paper would not change almost at all, and would not 
affect any conclusion, if the mean were not subtracted.}.

The choice of the normalization $M_{ti}=m_{ti}/x_{0i}$
to the initial price preserves the nature 
of the tails and is scale invariant. The use of the relative 
returns instead of the logarithm 
of the ratio of the consecutive returns, is motivated by
the additive rather than multiplicative character of the
price series. 

In the following, we will argue that the ${\bf C}$ covariance matrix
is able to reveal a fat-tail nature  
of the price change fluctuations, contrary 
to the covariances ${\bf G}$ and ${\bf J}$, 
defined below.  From the point of view of preserving 
power law tails, as a normalization
one could alternatively use the prices, $x_{\tau i}$,
at any random time $\tau$. However,  from the calculational
point of view involving the integrated return 
or the portfolio risk, the choice of the initial 
price is most convenient.

Commonly, the following covariance is used in the analysis:
\be
{\bf G}_{ij} = 
\frac 1T \,\sum_{t=1}^T \, \frac{m_{ti}}{\sigma_i}
\frac{m_{tj}}{\sigma_j}\,\,.
\label{G}
\ee
where now the normalization is given by the 
volatility (variance) 
$\sigma_i$ with $\sigma^2_i=\sum_t\,m_{ti}^2/T$.
In the presence of fat tails, {\em i.e.} if the 
distribution has a power-law behavior 
$p(\xi) \sim \xi^{-1-\alpha}$, $\alpha<2$, the
variance itself has a fat tail distribution with an index $\alpha/2$
and the average variance
does not exist. Obviously, in this case
the use of $\sigma$ as a normalization will
bias the analysis. 
Alternatively, in place of $\sigma_i$, one 
may use the quantity
$r_i = \sum_t |m_{ti}|/T$:
\be
{\bf J}_{ij} =
\frac 1T \,\sum_{t=1}^T \, 
\frac{m_{ti}}{r_i} \frac{m_{tj}}{r_j}\,\, .
\ee 
For $1<\alpha<2$, $r_i$ has a well defined  
large $T$ limit. However, in practice,
for finite $T$, this normalization, similarly as
the one for ${\bf G}$, obscures the effects 
of large price-change fluctuations.

\begin{figure}
\centerline{\includegraphics[width=0.48\textwidth]{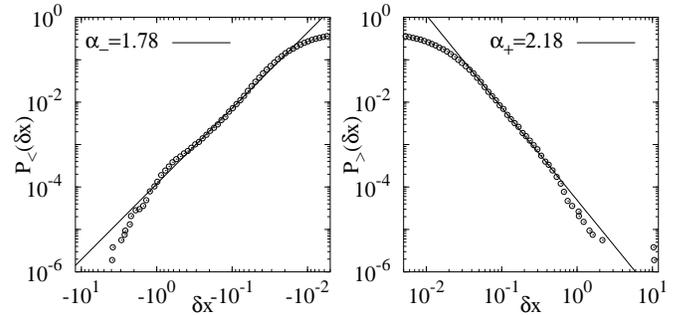}}
\bigskip
\caption{{\bf (a)} The cumulative distribution $P_<(\xi)$
for $M_{ti}=m_{ti}/x_{0t}$ as in ${\bf C}$ and the best fit
to the power law. {\bf (b)} The same for $P_>(\xi)$.}
\label{fig1}
\end{figure}

A common feature of the quantities
$M_{ti} = m_{ti}/x_{0i}$ for ${\bf C}$ and 
$M_{ti}=m_{ti}/\sigma_i$, $M_{ti}=m_{ti}/r_i$ 
for ${\bf G}$ and ${\bf J}$, 
is that they are invariant with respect to change
in the monetary unit. They exhibit, however, 
a different scale behavior for large $\xi$. 
The raw cumulative
probability $P_{<}(\xi)$ ($P_{>}(\xi)$)
defined as a probability that $M_{ti}$ 
is less (greater) than $\xi$, calculated for
all $i$ and $t$, is expected to pick up the 
smallest power of the tail, 
$\sim \pm A_\pm \xi^{-{\alpha_\pm}}$ (for gains/losses), 
present in the sample. In Fig.~\ref{fig1} we 
show the cumulative probabilities (a)
$P_{<}(\xi)$ and (b) $P_{>}(\xi)$ for ${\bf C}$.
The data in the figure are compared with the
power laws with: $\alpha_-=1.78$ and 
$A_-=2.4\cdot 10^{-3}$ and 
$\alpha_+=2.18$ $A_+=1.35\cdot 10^{-3}$. 
The values are given without errors.
The SP500 data set does not allow for an accurate 
determination of the fit parameters.
The given values have a qualitative meaning.
The fact that they are close to $2$ signals
the presence of fat tails of the underlying 
distribution. Indeed, the presence of fat-tails
will be confirmed by the following analysis of 
the covariance matrices. 

Repeating the same for ${\bf G}$ we find much
thinner tails with the following
exponents: $\alpha_-=3.8$ and $\alpha_+=4.5$,
and for ${\bf J}$: $\alpha_-=3.5$ and $\alpha_+=4.3$.

Clearly, the normalization to either the variance (volatility),
$\sigma_i$, or the range $r_i$ 
tends to affect the raw tail distributions, with a quenching 
towards the Gaussian distribution. This is expected, 
since the fluctuations are roughly normalized to the 
typical fluctuation. This is not the case in 
${\bf C}$, where the raw tail is retained. 

Let us introduce yet another 
covariance matrix which will be convenient in
the further analysis. We will construct it
from the signs $s_{ti} = {\rm sgn} \, m_{ti}$:
\be
{\bf S}_{ij} = 
\frac 1T \,\sum_{t=1}^T \, s_{ti}s_{tj}
\label{S}
\ee
We use in our analysis a three-valued sign function: 
${\rm sgn} = -1,0,1$. For all assets the average 
\be
\langle s_i \rangle = \frac 1T \,\sum_{t=1}^T s_{ti}
\approx 0
\ee
and the successive entries in the historically ordered row $s_{ti}$ are
essentially uncorrelated.

\section{Correlations}

By construction, the financial covariance matrix is composed of 
intra-assets (here stocks) correlations, and therefore tells us
how closely assets move in time-evolving market. The microscopic
nature of these correlations is so far unknown. However, a 
quantitative understanding can still be achieved statistically.
The source of the
correlations is two-fold: real correlations between assets
and statistical fluctuations. The statistical fluctuations
disappear in the $T=\infty$ limit. For finite $T$, however,
even in the absence of any real correlations, the non-diagonal
entries are non-zero. In the gaussian universality they fall-off
as $1/\sqrt{T}$. 

As a measure of correlations between distinct 
companies $i\ne j$ one can use a distribution $\rho({\bf C}_{ij})$.
Similar distributions can be constructed for ${\bf G}$ and ${\bf S}$.

\begin{figure}
\centerline{\includegraphics[width=0.48\textwidth]{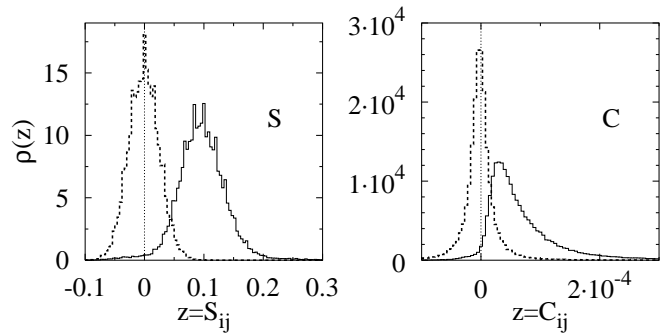}}
\bigskip
\caption{{\bf (a)} The distribution of correlations of signs
${\bf S}$ before (solid line) and after reshuffling (dashed line).
{\bf (b)} The same for ${\bf C}$.}
\label{fig7}
\end{figure}

Consider first correlations of pure signs.
In Fig.~\ref{fig7}.a, we show the distribution
(solid line) of ${\bf S}_{ij}$'s 
histogrammed over all pairs $i\ne j$. 
We observe a strong asymmetry
towards positive correlations
with a maximum around $0.1$,
indicating that assets have a tendency to move 
collectively in the same trend: up or down. 

This pronounced asymmetry is present in 
correlations for all 
other covariances ${\bf C}$, ${\bf J}$, ${\bf G}$.
We show in Fig.~\ref{fig7}.b 
the distribution of correlations for ${\bf C}$.
The correlations between signs are inherited
by all other covariances. We shall discuss 
possible consequences of this behavior 
in the section VI. The inter-stock (inter-sign)
correlations can be easily destroyed 
by a procedure of reshuffling 
described below. Indeed, we see in Fig.~\ref{fig7} 
(dashed line) that after reshuffling the spectra 
become symmetric.
 
\subsection{Reshuffling}

Let us introduce the abovementioned procedure to remove the inter-stock 
correlations from the data. Having done this, we will be
able to concentrate on the issue of the stochastic nature
of the price fluctuations.

The price changes $\delta x_i(t)$ enter the covariance
matrix in the historical order. This order, in particular
preserves inter-stock correlations. 
We can suppress the inter-stock correlations
in the data, by introducing a random 
time ordering to the time history for each asset. More precisely,
for each asset, $i$, we can generate a random permutation
of $t$-indices $P^{(i)}: t \rightarrow t' = P^{(i)}(t)$.
and instead of the historical ordered rows of returns
we can use: $\delta x'_i(t) = \delta x_i(t')$,
to define: $m'_{ti}= {\delta x'}_i(t) -{\delta \bar{x}}_i$ 
and the corresponding covariance matrices ${\bf C'}$, 
${\bf J'}$ ${\bf G'}$, and ${\bf S'}$. 
The permutations for different rows,
$P^{(i)}$, $P^{(j)}$, are random and mutually independent.
Such a reshuffling does not change the information content
of the individual asset rows, because successive
entries in the historical ordered row, 
$\delta x_i(t)$ and $\delta x_i(t+1)$, 
are uncorrelated for typical time intervals on a market. 
Thus, the reshuffling affects only
the inter-row information content
destroying any correlations. Hence we expect that 
the reshuffled data set should reflect pure
stochastic nature. Indeed, it does. 
In Fig.~\ref{fig7} we show for example
the effect of reshuffling on 
the inter-stock correlations. 
The asymmetry disappears.

\section{Spectra}

In this section we discuss the spectral density 
associated with the covariance matrices 
defined above. The spectral density
plays an important role in risk assessment~\cite{BOOK,RAN}.

\begin{figure}
\centerline{\includegraphics[width=0.48\textwidth]{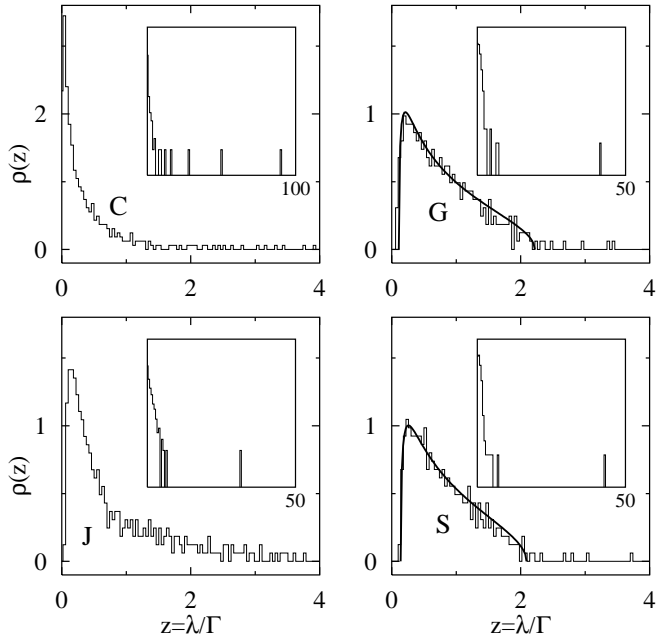}}
\bigskip
\caption{\label{fig2} The eigenvalue histograms
for ${\bf C}$,${\bf G}$,${\bf J}$ and ${\bf S}$. 
For ${\bf G}$ and ${\bf S}$ we present also the
gaussian fit~(\protect\ref{fitgauss}).
In the inlets we show the same histograms but
in a range embracing all eigenvalues. 
To make the smallest picks visible we 
artificially enhanced them by setting in inlets 
the logarithmic scale on the vertical axis. Notice
that except the one for ${\bf C}$ all plots have
the same ranges.}
\end{figure}

The results will be presented as histograms of eigenvalues 
$\lambda$. We will sort of unify 
the scale on the $\lambda$-axis by plotting
histograms as a function of a quantity 
$\lambda/\Gamma$ where $\Gamma$ is defined as:
\be
\Gamma = \frac{1}{N} \Tr {\bf C} = 
\frac{1}{NT} \sum_{ti} M^2_{ti} \, ,
\ee
and analogously for ${\bf G}$, ${\bf J}$ and ${\bf S}$. 
For a covariance of gaussian numbers,
the constant $\Gamma$ approaches a $T$-independent
constant in the limit $T\rightarrow \infty$ 
while for a covariance of power-law distributed numbers 
with $1<\alpha<2$, it behaves as 
$\Gamma = \gamma T^{2/\alpha-1}$, where $\gamma$
is a $T$-independent constant.
We will explain this scaling in more detail
in the section about random L\'evy matrices. 
It is easy to see that by
construction $\Gamma=1$ for $G$ and $S$. 
The histograms of eigenvalues $\lambda/\Gamma$ 
for the SP500 data are presented in Fig.~\ref{fig2}. 
Let us make here two points: the histograms
for ${\bf G}$ and ${\bf S}$ are almost identical.
All spectra have a few large eigenvalues. In the
next section using Random Gaussian Matrices (RGM)
we will define the scale which
will tell us which eigenvalues can be treated as large.

There are potentially two sources of large
eigenvalues in the spectrum: inter-stock correlations
and fat-tails. In the next sections we will discuss
methods to pinpoint the two effects.

\begin{figure}
\centerline{\includegraphics[width=0.48\textwidth]{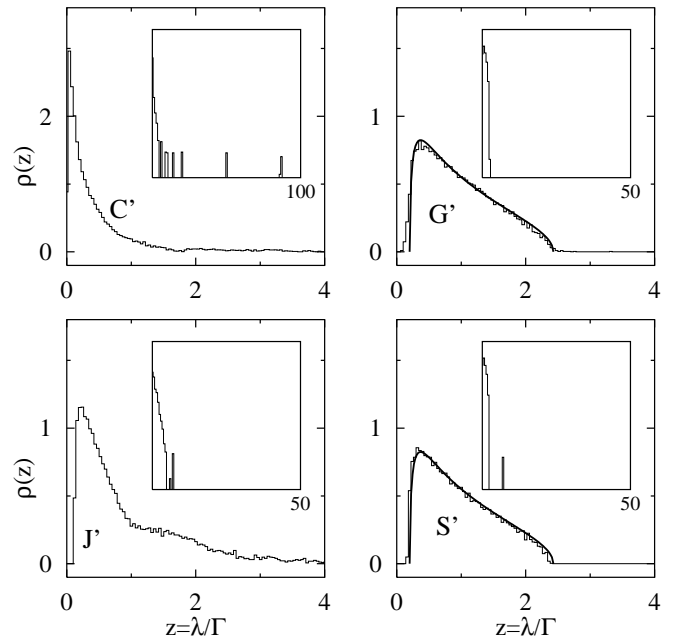}}
\bigskip
\caption{\label{fig3} The same as in fig.~\protect\ref{fig2} but 
for reshuffled data set. The histograms are averaged over
$20$ random reshufflings.}
\end{figure}

In the right block of Fig.~\ref{fig3} we plot the eigenvalue 
density of the
${\bf G'}$ and ${\bf S'}$ covariances. The spectra
for ${\bf G'}$ and ${\bf S'}$ are again almost identical.
In comparison with the spectra for the historically ordered
data set, right block of Fig.~\ref{fig2}, we see that the 
large eigenvalues 
disappear: the largest eigenvalue of ${\bf G}$ was 41.95,
and for ${\bf G'}$ after reshuffling 2.77. 
Moreover, the {\bf G'} and ${\bf S'}$ spectra fit very 
well to the curve:
\be
\rho(\lambda) 
\sim \frac 1\lambda 
\sqrt{(\lambda - \lambda_{min})(\lambda_{max} - \lambda)}
\label{fitgauss}
\ee
with $\lambda_{min}=0.20$, $\lambda_{max}=2.43$,
predicted by Random Gaussian Matrices (RGM)   
in the large $N$ limit~\cite{GAUSSQ}. In general
for the asymmetry parameter $a=T/N$, the formula
predicts $\lambda_{max,min}=(1\pm1/\sqrt{a})^2$,
which in particular for $a=T/N=1308/406\approx 3.22$, 
yields the values given above. These
values are used in the curve plotted in Fig.~\ref{fig3}.
This agreement clearly indicates that the normalization 
to the volatility $\sigma_i$ used in ${\bf G}$ brings 
the signal to the Gaussian universality. 

It is worth noting that if one attempts
to fit the RGM result to the eigenvalue histograms
for a correlated data set one generally obtains values for 
$\lambda_{min}$, $\lambda_{max}$ 
which deviate from the predicted ones. 
For example, for ${\bf G}$ and ${\bf S}$ for 
the historically ordered we get 
$\lambda_{max}=2.22$, $\lambda_{min}=0.11$
and $\lambda_{max}=2.09$, $\lambda_{min}=0.14$, respectively,
This deviation comes as a compensation
for the appearance of large eigenvalues which lie far away
and corresponds to asymmetry parameters $a=2.5$ and 
$a=2.9$, respectively.

As shown in Fig.~\ref{fig3}, the large eigenvalues 
survive reshuffling in ${\bf C}$ covariances.
Moreover, the position of the larger eigenvalues
is relatively stable under reshuffling. In fact,
in each of $20$ random reshufflings used in the
plot, the three larger eigenvalues normalized by 
the scale $\Gamma$, always 
land in the same histogram bin of the size 0.015.

The fact that there are large eigenvalues in
the reshuffled spectrum can be attributed to the presence 
of heavy tails in the probability distributions
for price fluctuations. The behavior
of the spectra can be understood, as we will see
below, in terms of Random L\'evy Matrix (RLM) 
theory.
  
\subsection{${\bf C}$ versus RLM}

We want to compare eigenvalue spectra of ${\bf C}$
and ${\bf C'}$ with the ones obtained from an
ensemble of Random L\'{e}vy Matrices (RLM)~\cite{BOUCIZ}.
We generate L\'{e}vy matrices (RLM) as
follows. We choose $N\times T$ elements of a matrix
$M_{ti}$ as independent random numbers 
from a L\'evy distribution. We find the eigenvalues
of the matrix $C_{ij} = 1/T \sum_{t} M_{ti} M_{tj}$.
We can repeat this many times collecting the
eigenvalues in a common histogram.

Here we would like to mention, that there exists an alternative construction
of random matrix ensembles, based on the concept of free random variables~\cite{VOICULESCU}. We call this realization 
Free L\'{e}vy Matrices (FLM)~\cite{USFRV}.
Ensembles of FLM, are, contrary to RLM, invariant under rotations.
They are more easily tractable using analytical methods. On the other side,
due to the invariance of the measure, the eigenvectors do not show 
interesting correlations like in the case of RLM, which we discuss
in the next sections. 
In the rest of this paper, we do not discuss FLM, and we refer for comparison
between the FLM and RLM ensembles to~\cite{USII.5}.

As an illustration,
in our numerical experiment we generated
% an ensemble of $n=xxxx$ *** For cumulative one matrix
% was generated ! *** 
RLM with 
%an asymmetry $N/T^{2/\alpha}\approx 3$ adjusted to 
an asymmetry $a=T/N=3.22$ adjusted to 
the asymmetry of the SP500 data set being presently considered.
The choice of the asymmetry will become clear in the 
next subsection.
As the distribution for the matrix elements $M_{ti}$
we took a symmetric L\'evy distribution
with $\alpha=1.7$ close to the value emerging from the analysis
of the tail behaviour of the cumulative probability $P_<(\xi)$.  

\begin{figure}
\centerline{\includegraphics[width=0.48\textwidth]{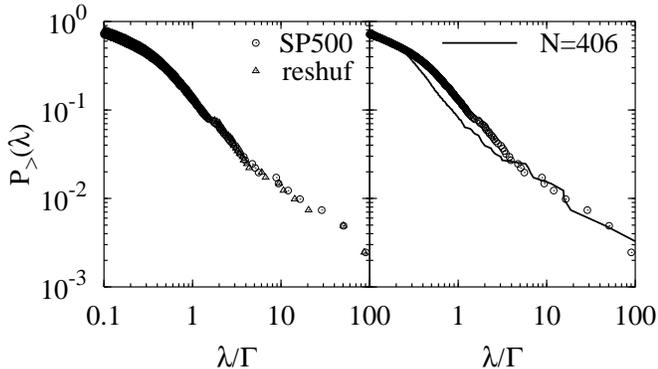}}
\bigskip
\caption{\label{fig4} {\bf (a)} Cumulative eigenvalue distribution for 
the SP500 data for the historically ordered data set compared
with the one for reshuffled data and
{\bf (b)} with the one of a randomly generated L\'evy matrix. }
\end{figure}

In Fig.~\ref{fig4} we compare the cumulative distributions
of eigenvalues for ${\bf C}$ and ${\bf C'}$, 
and the random matrix result for fixed
asymmetry $a=3.22$ and size $N=406$.
We see that the large eigenvalues 
in the spectrum ${\bf C}$ survive reshuffling.
The spectrum of RLM and of the reshuffled SP500 data set
exhibit similar large eigenvalue behaviour.

\subsection{Scaling in RLM}

Assume that $M_{ti}$ are power-law distributed: 
$p(\xi) \sim  \xi^{-1-\alpha}$. Define:
\begin{equation}
{\bf C}_{ij} = \frac{1}{T^\sigma} \sum_t M_{ti} M_{tj}
\end{equation}
with the normalization factor $1/T^\sigma$ whose exponent 
$\sigma$ may differ from $1$. We will argue that
a natural candidate for $\sigma$ is $\sigma=2/\alpha$.

Let us split ${\bf C}$ into a diagonal ${\bf D}$ 
and off-diagonal ${\bf A}$ parts
\begin{equation}
{\bf C}_{ij} = {\bf D}_{i} \delta_{ij} + {\bf A}_{ij}
\end{equation}
We can use the Central Limit Theorem for
L\'evy universality to obtain the distribution
of the entries in ${\bf C}$
in the large $T$ limit. We get
\begin{equation}
{\bf D}_{i} \sim T^{2/\alpha-\sigma} d_i \ , 
\quad {\bf A}_{ij} \sim T^{1/\alpha -\sigma} a_{ij} \, ,
\end{equation}
where $d_i$ and $a_{ij}$ are $T$ independent constants, distributed
with the stable L\'evy distributions. For the diagonal elements $d_i$ we
expect a distribution with an index $\alpha/2$ and the skewness
parameter $\beta=1$, while the off-diagonal elements $a_{ij}$ will
be distributed with a symmetric distribution with an index $\alpha$.
To assess the importance of the off-diagonal entries on the
spectrum, we use the standard perturbation theory. For that, we 
write
\begin{equation}
{\bf C}_{ij} = T^{2/\alpha-\sigma} c_{ij} =
T^{2/\alpha-\sigma} \left( d_{i} \delta_{ij} 
+ T^{-1/\alpha} a_{ij} \right) \ .
\end{equation}
and expand $c_{ij} = d_i \delta_{ij} + \epsilon a_{ij}$ 
in $\epsilon = 1/T^{1/\alpha}$.
In zeroth order, the eigenvalues of ${\bf C}_{ij}$ are just $d_i$.
The first order corrections are zero because
the matrix ${\bf A}_{ij}$ is off-diagonal. 
Generically, for a random matrix, $d_i$'s are not degenerate, 
so up to the  second order, the eigenvalues of ${\bf C}_{ij}$ are
\begin{equation}
\lambda_i = d_i + \epsilon^2 \sum_{j (\ne i)} \frac{a^2_{ij}}{d_j-d_i} 
= d_i + T^{-2/\alpha} \sum_{j (\ne i)} \frac{a^2_{ij}}{d_j-d_i} 
\end{equation}
There are $N-1$ terms in the sum, each of order unity. Thus the
sum contributes a factor proportional to $N$, say $\approx s_i N$,
and we have:
\begin{equation}
\lambda_i =  d_i + s_i N T^{-2/\alpha} \, .
\end{equation}
The off-diagonal terms compete with the diagonal ones for 
$N\approx T^{2/\alpha}$. In our case, $\alpha=1.7$, 
$N/T^{2/\alpha}\approx 1/3$. 
The range of the spectrum of ${\bf C}$ 
will not grow with $T$ for $\sigma=2/\alpha$.

The normalization constant 
$\Gamma = \langle \Tr {\bf C} \rangle /N$,
which we have introduced previously for the experimental
covariances, behaves for RLM with $1<\alpha<2$ as
$\Gamma = \gamma T^{2/\alpha-\sigma}$.
Again it simplifies for the choice
$\sigma=2/\alpha$.

\section{States}

In this section we analyze the eigenvector content of the three
covariances using the inverse participation ratios and the 
stock scattering. We show that the covariances with 
larger tails are more stable under reshuffling the SP500
data, with localized states at the edge of the spectrum.

\subsection{Inverse Participation Ratios}

To better understand the nature of large-eigenvalues
in the SP500 data, we now turn to the eigenvector content. 
For that we use the inverse participation
ratio
\begin{equation}
{\bf Y}_\lambda = \sum_{i=1}^N \,V_{\lambda i}^4
\label{add100}
\end{equation}
where $V_\lambda$ is a normalized~\footnote{This 
implicitly assumes that the entries  $V_{\lambda i}$ are 
at least power law distributed with index $\alpha>1$.}
eigenvector:
\begin{equation}
\sum_{i=1}^N \,V_{\lambda i}^2 = 1
\end{equation}
of ${\bf C}$ to the eigenvalue $\lambda$. 
We can  distinguish the `mixed' states with 
${\bf Y}_\lambda \approx 1/N \approx 0$ and
`pure' states with  ${\bf Y}_\lambda \approx 1$.  

\begin{figure}
\centerline{\includegraphics[width=0.48\textwidth]{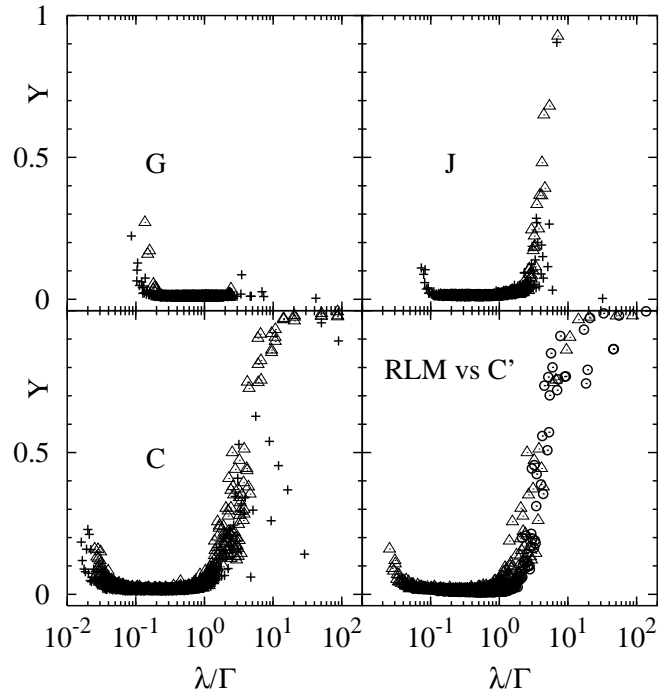}}
\bigskip
\caption{\label{fig5} Inverse participation ratios for
for the SP500 covariances, for the historically ordered data 
(crosses) and for the reshuffled data (triangles). In the
lower right figure we compare the distribution of the
reshuffled data (triangles) and of Random L\'evy Matrix (circles).}
\end{figure}

In Figs.~\ref{fig5} we display the inverse participation 
ratios for the three covariances ${\bf C}$, ${\bf G}$ 
and  ${\bf J}$ for the raw SP500 data 
(triangles) and reshuffled SP500 data (pluses).
Additionally, in the fourth insert we compare
the inverse participation ratio for ${\bf C'}$ (triangles)
and for RLM with index $\alpha=1.5$. 
The large eigenvalues 
are localized for ${\bf J}$ and  ${\bf C}$
with intermediate and large
participation ratio, respectively. 
For ${\bf G}$ and ${\bf J}$
the largest eigenvalue states are `mixed' 
while for ${\bf C}$ they are `pure'.
Large eigenvalue pure states are present
also in the ${\bf C'}$ covariance ! 
This is clearly displayed by the data. 
The inverse participation ratio as a function
of eigenvalue for ${\bf C'}$ has 
the same character as RLM.

The L\'evy randomness has an equally strong effect
on the large part of the distribution as the
inter-stock correlations. Also, 
in the ${\bf G}$ and ${\bf J}$ covariances
reshuffling `breaks' the clustering, removing 
from the spectrum large eigenvalues. 

Concerning the lower part of the spectrum,
it is interesting to note that it 
is characterized by smaller
participation ratios than the large eigenvalue part.
The exception is ${\bf G}$ for which the part with small 
eigenvalues has larger inverse participation ratio.
On the other hand, the low eigenvalue 
parts of the ${\bf G}$ and ${\bf C}$ show a similar
behavior. In fact, one expects
that the shape of the lower part of the spectra 
for RLM and RGM to be  strongly related to the asymmetry
of the matrices rather than to the type of randomness.
Indeed, for matrices with asymmetry $a<1$, the spectra
exhibit exact zero eigenvalue states (zero modes). 

\subsection{Stock scattering}

In analogy to the inverse participation ratio
we define a quantity:
\begin{equation}
{\bf P}_i = \sum_\lambda \,V_{\lambda i}^4
\end{equation}
which measures how many eigenstates are
mixed in a pure stock state $i$. 
We will refer to it as the stock scattering.
Again we have the normalization:
\begin{equation}
\sum_{\lambda} \,V_{\lambda i}^2 = 1
\end{equation}
The inverse $1/{\bf P}_i$ tells us how many eigenstates
are influenced by the stock $i$. 

\begin{figure}
\centerline{\includegraphics[width=0.48\textwidth]{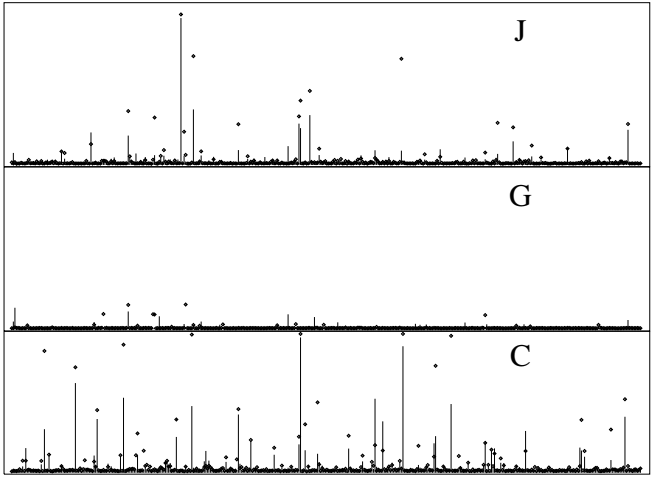}}
\bigskip
\caption{\label{fig6} Stock scattering for the SP500 covariances
for the historically ordered data (lines) and the reshuffled
data (dots).}
\label{figr}
\end{figure}

In Fig.~\ref{fig6} we show the value of ${\bf P}_i$ for the consecutive
406 stocks (horizontal) with $0\leq {\bf P}_i\leq 1$ (vertical),
for each of the three covariances discussed in this work, at the
same scale to facilitate the comparison. Clearly, the normalization
to unit volatility drives ${\bf P}_i$
towards gaussian noise. We have checked that the effects of reshuffling
is to enhance the localization of certain stocks (without much affecting
the original ones), in agreement with the participation ratio analysis.
The stock scattering of ${\bf C}$ provides
a relatively simple filter for those
stocks that localize in a market, and are likely 
to drive the large tail behavior of the covariance matrix. 

\section{Model} 

Collecting all experimental evidence for
the SP500 data we are led 
to the conjecture, that 
the returns normalized to the initial price 
$M_{ti} = m_{ti}/x_{0i}$ undergo fluctuations
which can be well described by a randomness 
of the type:
\begin{equation}
M_{ti} = {\rm sgn}_{ti} \cdot \xi_{it} \, 
\label{hip}
\end{equation}
where ${\rm sgn}_{ti}$ is a random matrix of correlated
signs and $\xi_{it}$ are identical independent
distributed numbers from the L\'evy universality. 
Indeed, as shown in Fig.~\ref{fig7} the sign correlations
are present in all covariances. Additionally, one can
check that the substitution of signs of $\delta x_i(t)$ by
random signs has the same effect on spectra of the SP500 covariances
as reshuffling. 

Moreover, the spectra of eigenvalues of ${\bf G}$ 
and ${\bf S}$ are almost identical and can be seen from
the Fig.~\ref{fig2} and from the comparison of a few largest
eigenvalues which are $6.89$, $7.27$ and $41.95$ for ${\bf G}$
and $5.48$, $7.43$ and $43.25$ for ${\bf S}$.
This tells us that the information about the 
inter-stock correlations present in ${\bf G}$ is 
already present in ${\bf S}$. Thus, as long as the long 
tails are suppressed, as in ${\bf G}$, 
the absolute value of the fluctuations 
does not matter. 

The absolute value of the changes matters, however, if we do 
not introduce any superfluous normalizations 
and expose the covariance\footnote{One can define 
other covariances, like for example, the one constructed 
from instantaneous returns: $M_{ti} = m_{ti}/x_{ti}$,
or $M_{ti} = \log \delta x_i(t+1)/\delta x_i(t)$, 
which similarly to ${\bf C}$,
would preserve information about the power-law tails.
As a consequence, for instance, 
their spectra would behave in the same way 
under the reshuffling as the one of ${\bf C}$. }
to large price fluctuations as in ${\bf C}$.
In this case, the correlations of 
signs play a secondary role, since even 
when one decorrelates them by reshuffling, 
the large eigenvalues stay in the spectrum.

Similarly as under reshuffling, the spectrum 
of ${\bf C}$ does not change its character and the
largest eigenvalues are quite stable, if one uses
randomly generated signs insteads of those inherited
from the historical data.

The randomness given by the 
formula (\ref{hip}) captures many experimentally 
observed features of the real data. It should
be treated, however, as as zeroth order approximation.
In a more involved analysis, one should 
introduce some corrections to the conjecture 
which for example can take into account the possibility
of correlations between sign and absolute value of 
the price changes. Indeed, the data show the
existence of such correlations. 

\section{Conclusions}

We have carried an empirical analysis of the covariance 
matrices characterizing daily price returns from the SP500 market. 
We have shown that a specific covariance (returns normalized to the
stock initial price) exhibits matrix entries with almost stable
L\'{e}vy tails. A comparative study shows that only this covariance 
is stable under reshuffling, with a spectrum in remarkable agreement
with the one extracted from an ensemble of random L\'{e}vy matrices
with commensurate sizes and asymmetry. An analysis of the corresponding
participation ratio shows large localized and almost `pure' states.
This is not the case of the other covariances (returns normalized to
the stock mean mean volatility or range), which are characterized
by `mixed' states with one characteristically large and delocalized 
eigenvalue reminiscent of Yang's ODLRO~\cite{YANG}. The stock content of the
localized states is best displayed using the stock scattering.

In nearly gaussian markets, the risk is usually assessed
by minimizing the variance of a pertinent market policy,
say an investment portfolio, using the empirical
market covariance as suggested by Markowitz~\cite{MARKOWITZ}. Recently,
it was pointed out that the low-lying
eigenvalues of the empirical market covariance are gaussian
noise dominated (information free), implying that standard 
Markowitz's theory for risk assessment is flawed~\cite{RAN}.
In non-gaussian markets, the potential for large asset 
fluctuations may require using an alternative to Markowitz's
theory through the use of value-at-risk or tail-covariance
\cite{BOOK}, each of which requiring the covariance matrix.

In the present work we have
shown that eigenvalues of a market covariance 
follow the theoretical distribution of eigenvalues
of almost randomly generated L\'{e}vy matrices. The empirical market
covariance reflects on a state of maximum entropy in the
generalized sense of Dyson for random L\'{e}vy matrices.
Our observations maybe relevant for assets diversification and
risk management.

\vskip 2cm
{\bf Acknowledgments:}
\vskip .5cm
This work was supported in part by the US DOE grant DE-FG02-88ER40388,
by the Polish Government Project (KBN)  2P03B 01917,
by the Hungarian FKFP grant 220/2000, and by
the EC IHP grant HPRN-CT-1999-00161.


\begin{thebibliography}{33}


\bibitem{BOOK}
R.~Mantegna and H.~Stanley, 
{\it An Introduction to Econophysics}, Cambridge Univ. (2000); 
J.Bouchaud and M. Potters, {\it Theory of Financial Risks}, Cambridge Univ.
 (2000).
 
\bibitem{BOOKLEV}
For a  review, see e.g. 
{\it L\'{e}vy Flights and Related Topics},
Eds. M. Shlesinger, G. Zaslavsky and U. Frisch,
Springer 1995.



\bibitem{QCD}
R.A. Janik, M.A. Nowak, G. Papp and I. Zahed, 
{\it Acta Phys. Pol.} {\bf B28} (1997) 2949 {\tt e-print hep-th/9710103}.


\bibitem{RAN}
L.~Laloux, P.~Cizeau, J.~Bouchaud and M.~Potters,
{\it Phys. Rev. Lett.} {\bf 83}  (1999) 1467, {\tt e-print cond-mat/9810255};
V.~Plerou, P.~Gopikrishnan, B.~Rosenow, L.~Nunes~Amaral, and H.~Stanley,
{\it Phys. Rev. Lett.} {\bf 83} (1999) 1471, {\tt e-print cond-mat/9902283}.
For very recent applications, see e.g. P.~Gopikrishnan, B.~Rosenow, V.~Plerou
and H.E.~Stanley,
{\it Identifying Business Sectors from Stock Price
Fluctuations}, {\tt e-print} cond-mat/0011145, S. Dro\.{z}d\.{z}, 
J. Kwapie\'{n}, F. Gruemmer, F.Ruf and J. Speth, 
{\it Quantifying dynamics of the financial correlations},
{\tt e-print} cond-mat/0102402 and references therein. 

\bibitem{USFRV}
Z. Burda, R.A. Janik, J. Jurkiewicz, M.A. Nowak, G. Papp and I. Zahed,
{\it Free Random L\'{e}vy Matrices}, {\tt e-print} cond-mat/0011451. 


\bibitem{USII.5}
Z. Burda, J. Jurkiewicz, M.A. Nowak, G. Papp and I. Zahed,
{\it Free Random L\'{e}vy Matrices and Financial Correlations},
{\tt e-print} cond-mat/0103109.

\bibitem{GAUSSQ}
This formula was obtained  by several authors using various techniques, 
see e.g. A. Crisanti and H. Sompolinsky, Phys. Rev. {\bf A36} (1987) 4922;
A. Edelman, SIAM J. Matrix Anal. Appl. {\bf 9} (1988) 543;
M. Opper, Europhys. Lett. {\bf 8} (1989) 389;
J. Feinberg and A. Zee, {\it Renormalizing Rectangles and Other Topics
in Random Matrix Theory}, {\tt e-print} cond-mat/9609190;
R.A. Janik, M.A. Nowak, G. Papp, J. Wambach and I. Zahed, 
Phys. Rev. {\bf E55} (1997) 4100;
A.M. Sengupta and P.P. Mitra, Phys. Rev. {\bf E60} (1999) 3389.
\bibitem{BOUCIZ}
P. Cizeau and J.P. Bouchaud, {\it Phys. Rev.} {\bf E50} (1994) 1810. 



\bibitem{VOICULESCU}
D.V. Voiculescu, {\it Invent. Math.} {\bf 104} (1991) 201;
D.V. Voiculescu, K.J. Dykema and A. Nica, ``Free Random Variables'',
 Am. Math. Soc., Providence, RI (1992); for new results see also
A. Nica and R. Speicher, {\it Amer. J. Math.} {\bf 118} (1996) 799;
H. Bercovici and D. Voiculescu, {\it Ind. Univ. Math. J.}{\bf 42} (1993) 733.

%\bibitem{USII.75}
%Z. Burda, J. Jurkiewicz, M.A. Nowak, G. Papp and I. Zahed,
%{\it Free Random L\'{e}vy Covariances}.

\bibitem{YANG}
C.N.~Yang, Rev. Mod. Phys. {\bf 34} (1962) 694.

\bibitem{MARKOWITZ}
H. Markowitz, {\it Portfolio Selection: Efficient Diversification of 
Investments}, J. Wiley and Sons, New York (1959).
\end{thebibliography}
\end{document}